\documentclass[prapplied,superscriptaddress,aps,amsmath,amssymb,floatfix,reprint,raggedbottom]{revtex4-1}
\usepackage{graphicx}
\usepackage[colorlinks=true,citecolor=blue,linkcolor=blue,urlcolor=blue]{hyperref}
\usepackage{amsmath}
\usepackage{subfigure}
\usepackage{dcolumn}
\usepackage{xcolor}
\usepackage{fancyhdr}
\usepackage{lipsum}
\usepackage{soul}
\usepackage[english]{babel}
\usepackage{footnote}
\usepackage{fancyhdr}
\makeatletter 
\renewcommand{\fnum@figure}{\textbf{FIG.~\thefigure}}
\makeatother
\makeatletter
\def\bbordermatrix#1{\begingroup \m@th
  \@tempdima 4.75\p@
  \setbox\z@\vbox{%
    \def\cr{\crcr\noalign{\kern2\p@\global\let\cr\endline}}%
    \ialign{$##$\hfil\kern2\p@\kern\@tempdima&\thinspace\hfil$##$\hfil
      &&\quad\hfil$##$\hfil\crcr
      \omit\strut\hfil\crcr\noalign{\kern-\baselineskip}%
      #1\crcr\omit\strut\cr}}%
  \setbox\tw@\vbox{\unvcopy\z@\global\setbox\@ne\lastbox}%
  \setbox\tw@\hbox{\unhbox\@ne\unskip\global\setbox\@ne\lastbox}%
  \setbox\tw@\hbox{$\kern\wd\@ne\kern-\@tempdima\left[\kern-\wd\@ne
    \global\setbox\@ne\vbox{\box\@ne\kern2\p@}%
    \vcenter{\kern-\ht\@ne\unvbox\z@\kern-\baselineskip}\,\right]$}%
  \null\;\vbox{\kern\ht\@ne\box\tw@}\endgroup}
\makeatother

\usepackage{letltxmacro}

\LetLtxMacro{\ORIGselectlanguage}{\selectlanguage}
\makeatletter
\DeclareRobustCommand{\selectlanguage}[1]{%
  \@ifundefined{alias@\string#1}
    {\ORIGselectlanguage{#1}}
    {\begingroup\edef\x{\endgroup
       \noexpand\ORIGselectlanguage{\@nameuse{alias@#1}}}\x}%
}
\newcommand{\definelanguagealias}[2]{%
  \@namedef{alias@#1}{#2}%
}
\makeatother

\definelanguagealias{en}{english}

\begin{document}
\title{Subnanosecond Fluctuations in Low-Barrier Nanomagnets}
\author{J. Kaiser}
\author{A. Rustagi}
\author{K. Y. Camsari}
\affiliation{School of Electrical and Computer Engineering, Purdue University, IN, 47907}
\author{J. Z. Sun}
\affiliation{IBM Research Division, Thomas J. Watson Research Center, P.O. Box 218, Yorktown Heights, NY 10598}
\author{S. Datta}
\author{P. Upadhyaya}
\affiliation{School of Electrical and Computer Engineering, Purdue University, IN, 47907}

\date{\today}

\begin{abstract}
Fast magnetic fluctuations due to thermal torques have useful technological functionality ranging from cryptography to probabilistic computing. The characteristic time of fluctuations in typical uniaxial anisotropy magnets studied so far is bounded from below by the well-known energy relaxation mechanism. This time scales as $\alpha^{-1}$, where $\alpha$ parameterizes the strength of dissipative processes. Here, we theoretically analyze the fluctuating dynamics in easy-plane and antiferromagnetically coupled nanomagnets. We find in such magnets, the dynamics are strongly influenced by fluctuating intrinsic fields, which give rise to an additional dephasing-type mechanism for washing out correlations. In particular, we establish two time scales for characterizing fluctuations (i) the average time for a nanomagnet to reverse|which for the experimentally relevant regime of low damping is governed primarily by dephasing and becomes independent of $\alpha$, (ii) the time scale for memory loss of a single nanomagnet|which scales as $\alpha^{-1/3}$ and is governed by a combination of energy dissipation and dephasing mechanism. For typical experimentally accessible values of intrinsic fields, the resultant thermal-fluctuation rate is increased by multiple orders of magnitude when compared with the bound set solely by the energy relaxation mechanism in uniaxial magnets. This could lead to higher operating speeds of emerging devices exploiting magnetic fluctuations.

\end{abstract}
\maketitle
\thispagestyle{fancy}

\section{INTRODUCTION}
Nanoscale magnets driven by torques due to magnetic fields \cite {tehrani_progress_1999}, charge currents \cite{slonczewski_current-driven_1996,*berger_emission_1996,*brataas_current-induced_2012}, electric fields \cite{ohno_electric-field_2000,*wang_effect_2014}
and thermal fluctuations \cite{brown_thermal_1963,brown_thermal_1979,coffey_thermal_2012} have attracted rigorous interest in the recent past. On a fundamental level, these nanomagnets have served as a model dynamical system for studying the interplay of nonlinear dynamics and stochastic processes \cite{mayergoyz_nonlinear_2009}. While on the technological front, nanomagnets are being considered as promising next-generation memory \cite{diao_spin-transfer_2007,*chen_advances_2010}, communication\cite{locatelli_spin-torque_2014}, and information processing elements \cite{niemier_nanomagnet_2011,*csaba_nanocomputing_2002}. Traditionally, such applications require encoding information in the stable configurations of the magnetic order parameter. Consequently, both fundamental and technological studies have primarily focused on the regime when the energy barrier between the stable states of the magnet is much larger than the thermal energy, referred to as the nonvolatile regime. More recently, it has been realized that the order-parameter dynamics even in the other extreme, namely the low-barrier volatile regime, can be utilized to engender useful technological functionality, including true random-number generation \cite{vodenicarevic_low-energy_2017,*parks_superparamagnetic_2018,*debashis_tunable_2018}, probabilistic computing \cite{camsari_stochastic_2017,faria_low-barrier_2017,borders_integer_2019}, optimization \cite{sutton_intrinsic_2017}, machine learning \cite{zand_low-energy_2018,*pyle_hybrid_2018} and quantum emulation \cite{camsari_scalable_2019}. 
\begin{figure}[b]
{\includegraphics[trim={10 15 5 5},clip,width=1\linewidth]{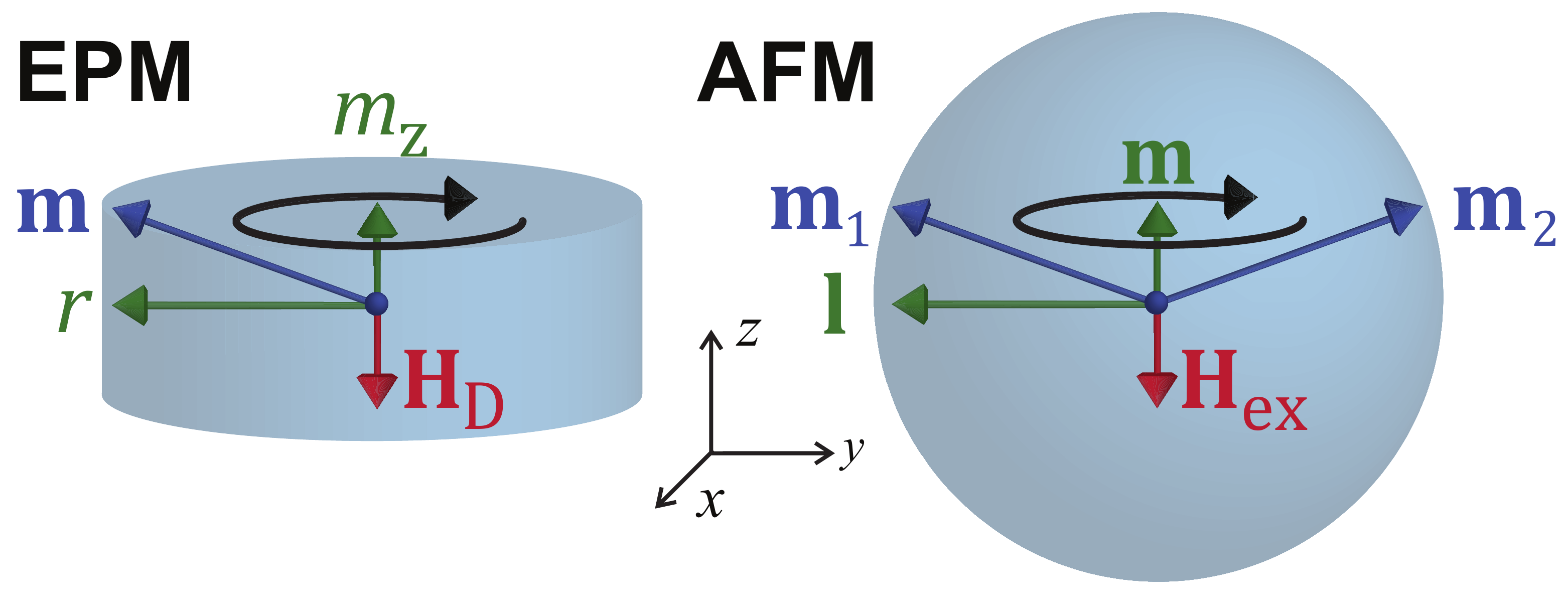}}
\caption{Illustration of magnet dynamics of EPM and AFM in the low $\alpha$ regime. The strong intrinsic field speeds up the overall dynamics. The demagnetization field is given by  $\mathbf{H}_D=-H_D m_z \mathbf{z}$ for EPM and the exchange field of AFM by $\mathbf{H}_\mathrm{ex}=-H_\mathrm{ex} \mathbf{m}$.}
\label{fig: fig1}
\end{figure}

The dynamics in nanomagnets are dependent on thermal fluctuations which excite the order parameter. The time it takes for a magnet to lose its memory can be characterized by its correlation time. Decreasing the correlation time results in an increase of operating speed for emerging applications in the low-barrier volatile regime. In the high-barrier regime, the correlation time is governed by energy relaxation and can be described by an Arrhenius relation of the form $\tau_c \propto \tau_0 \exp\left(\Delta/\left(k_B T\right)\right)$ with $\tau_0=1/(\alpha \gamma H_K)$ where $\alpha$ is the Gilbert damping constant, $\gamma$ is the gyromagnetic ratio, $H_K$ is the uniaxial anisotropy, $\Delta$ the barrier of the magnet, $k_B$ the Boltzmann constant and $T$ the temperature \footnote{CGS units are used throughout the paper}. This formula suggests that it is impossible to decrease $\tau_c$ by decreasing $H_K$ to very small values since $\Delta$ scales as $H_K$ but $\tau_0$ scales as $1/H_K$. However, when the barrier becomes comparable to the thermal energy $k_B T$, the Arrhenius formula does not apply. In his seminal paper, Brown derived an expression for the relaxation time for magnets with uniaxial anisotropy in the low-barrier approximation which is scales as $M_s V/(\alpha \gamma k_B T)$ for $\Delta \rightarrow 0$ where $M_s$ is the saturation magnetization and $V$ is the volume \cite{brown_thermal_1963,coffey_thermal_2012}. This formula characterizes the fundamental limit for decreasing the time scales of fluctuations of a magnet with uniaxial anisotropy by decreasing its barrier.

In this Paper, we analyze the magnetization dynamics due to thermal excitation for easy-plane (EPM) and antiferromagnetically coupled low-barrier nanomagnets (AFM) in the limit $\Delta \rightarrow 0$. 
For both systems shown in Fig. \ref{fig: fig1}, we answer the questions how fast the magnetization changes on average in thermal equilibrium (parametrized by $\tau_r$) and how fast the magnetization loses its memory (parameterized by $\tau_c$). As the first main result, we show that  
\begin{equation}
\tau_r \propto \frac{1}{\gamma \sqrt{H_\mathrm{in} H_\mathrm{th} }},
\label{eqn: eqn_tau_r}
\end{equation}
where $H_\mathrm{th} \equiv k_B T/(M_s V)$ and $H_\mathrm{in}$ is the intrinsic field \footnote{$H_{\rm th}$ is the field in magnitude corresponding to thermal energy for a given magnetic body of moment ($M_s V$), and is different from the Langevin field $H_{\rm fl}$ used for finite temperature stochastic LLG equation under thermal agitation as defined in Eqn. \ref{eqn: Hfl}}. In particular, we highlight that this time is independent of the damping parameter $\alpha$. The intrinsic field is the demagnetization field $H_D$ for EPM and the exchange field $H_\mathrm{ex}$ for AFM. In this case, the order parameter changes on average due to different precession frequencies (which is referred to here as the dephasing mechanism) caused by the thermal fluctuations in equilibrium.
As the second main result we find that, due to the presence of dephasing in addition to energy relaxation, the time for memory loss is described by
\begin{equation}
\tau_c \propto \frac{1}{\alpha^{1/3} \gamma H_\mathrm{in}^{2/3} H_\mathrm{th}^{1/3}}.
\label{eqn: eqn_tau_ensemble}
\end{equation}
In the experimentally relevant parameter regime, due to the strong intrinsic field, the time for memory loss is in the subnanosecond regime and orders of magnitude smaller than the fundamental limit for fluctuations of uniaxial anisotropy magnets. This fast memory loss can increase the operating speed of emerging devices.

\section{MAGNETIZATION DYNAMICS}
Within the single domain limit of stochastic Landau-Lifshitz-Gilbert (sLLG) phenomenology \cite{berkov_magnetization_2007}, the order-parameter dynamics for nanomagnets are governed by $\mathbf{\dot{m}}_i=- \gamma \mathbf{m}_i \times\mathbf{H}_\mathrm{eff,i} + \alpha \mathbf{m}_i \times \mathbf{\dot{m}}_i$, where $\gamma$ is the gyromagnetic ratio, and $\alpha$ is the Gilbert damping parameter. $\mathbf{H}_\mathrm{eff,i}=-\delta \mathcal{F}/(\delta {\bf m}_i M_s) + {\bf H_{\rm fl}}$ is the effective magnetic field, where the first term describes contributions from external and internal fields derived from a free-energy density $\mathcal{F}$, and the second term denotes fields due to thermal fluctuations \cite{berkov_magnetization_2007}.

For EPMs, $\mathbf{{m}}_i = {\bf m} \equiv (m_x,m_y,m_z)$, with $\mathbf{m}$ being the unit-vector order parameter oriented along the magnetization, while $\mathcal{F}_\mathrm{EPM}=H_DM_sm_z^2/2-H_KM_sm_x^2/2$. The first term in the free energy represents the out-of-easy-plane demagnetization energy, with $H_D=4\pi M_s$, while the second term denotes anisotropy energy due to a uniaxial anisotropy field $H_K$. The minimum energy configurations are obtained when the magnetization lies within a plane (easy plane). Such EPMs are naturally formed in thin-film circular ferromagnets, where the shape-induced dipolar energy defines an easy plane to be normal to the thickness \cite{cowburn_single-domain_1999,*leo_collective_2018}. The case with an $x-y$ easy plane is shown in Fig. \ref{fig: fig1}. Thermal fields in this case give rise to random deviations, $m_z$, away from the easy plane. This results in fluctuating internal fields $H_\mathrm{in}=H_Dm_z$ oriented normal to the easy plane, and causes precession of ${\bf m}$ around it.

AFMs consist of two negatively exchange coupled magnetic sublattices, which can either occur naturally \cite{baltz_antiferromagnetic_2018} or be synthesized by coupling two ferromagnets via negative Ruderman-Kittel-Kasuya-Yosida (RKKY) interactions \cite{duine_synthetic_2018}. The order parameter in AFM is parametrized by the N\'eel vector ${\bf l}\equiv ({\bf m_1} -{\bf m_2})/2$, where ${\bf m_1}$ and ${\bf m_2}$ are unit vectors oriented along the sublattice magnetizations. For AFMs, ${\bf m_i}$ labels the sublattice magnetization unit vector, which is related to the N\'eel order parameter by ${\bf l}\equiv ({\bf m}_1 -{\bf m}_2)/2$, and  $\mathcal{F}_\mathrm{AFM}=H_{\rm ex} M_s {\bf m}_i \cdot {\bf m}_j-H_K M_s (m_\mathrm{1,x}^2+m_\mathrm{2,x}^2)/2$. In the absence of thermal fields, the requirement to minimize the exchange energy enforces the configuration with antialigned sublattice magnetizations, with ${\bf m} \equiv ({\bf m_1} +{\bf m_2})/2 =0$. Thermal fields disturb this configuration by canting sublattice magnetizations and produce a random nonzero ${\bf m}$. As shown in Fig. \ref{fig: fig1} similar to EPMs, this deviation is accompanied by generation of a random internal field $H_\mathrm{in}=H_{\rm ex} |\mathbf{m}|$ normal to the order parameter ${\bf l}$, and gives rise to a precession of ${\bf l}$. 

We are specifically interested in the regime where $M_s V H_K/(2k_BT) \rightarrow 0$. Experimentally such an AFM system can be built as a synthetic antiferromagnet (SAF) \cite{zhang_l10fe_2018} where two low-barrier ferromagnets are negatively coupled through RKKY interaction. These superparamagnetic magnets can be realized in a magnetic system where the surface anisotropy counteracts the shape anisotropy \cite{vodenicarevic_low-energy_2017, *parks_superparamagnetic_2018, *debashis_tunable_2018, vodenicarevic_circuit-level_2018, *mizrahi_neural-like_2018,*zink_telegraphic_2018,*debashis_experimental_2016}. In this low-barrier regime, the strength of internal fields for EPM and AFM are dominated by $H_D$ and $H_{\rm ex}$, respectively.

To characterize the magnetic fluctuations of EPM and AFM, we analyze their correlation in a particular direction (here in $+x$ direction). The correlation function is defined by $C(t)=\langle \mathcal{O}(0)\mathcal{O}(t)\rangle$, where $\mathcal{O}=m_x$ and $\mathcal{O}=l_x$  for EPM and AFM, respectively, and the brackets $\langle ... \rangle$ denote ensemble average. We derive and analyze the correlation function and time scales for two distinct cases. First, we derive the average reversal time $\tau_r$, defined as the characteristic time in which the correlation function changes in thermal equilibrium.  Then, we derive the memory loss time $\tau_c$, defined as the characteristic time in which the correlation function changes for an identically initialized ensemble of nanomagnets. In this case the system is out of thermal equilibrium.

\section{MAGNETIZATION REVERSAL}
To analytically understand the influence of internal fields on magnetization reversal, we focus on the regime where the internal fields dominate the dynamics of EPM and AFM. To this end, for analytical models we consider $H_K=0$ and $H_D$, $H_{\rm ex} \gg H_{\rm th}$. 
For EPM, due to large $H_D$, the magnetization excursions out of the easy $x-y$ plane are small, that is $m_z \ll 1$. The deterministic dynamics of the order parameter, as derived by expanding LLG up to the first order in $m_z$, can then be expressed in cylindrical coordinates $(m_x,m_y,m_z)\rightarrow(r,\varphi,m_z)$ as \cite{takei_superfluid_2014}:
\begin{subequations}
\begin{align}
\dot{\varphi}+\alpha \dot{m}_z =- \gamma H_D m_z,
\\
\dot{m}_z-\alpha \dot{\varphi} =0.
\end{align}
\label{eqn: dphidt}
\end{subequations}
Next, we utilize the fact that experimentally relevant systems typically have low Gilbert damping. In this limit (we derive the validity range of $\alpha$ \textit{a posteriori}), we can write Eqn. \ref{eqn: dphidt} as $\dot{\varphi} =- \gamma H_D m_z$ and $\dot{m}_z \approx 0$. The resultant dynamics can be understood as the precession of the order parameter around the demagnetization field ${\bf H}_D=-H_Dm_z {\bf z}$. The time-dependent angle $\varphi(t)$ of the in-plane precession is then given by
\begin{equation}
\varphi_{\rm EPM}(t) =- \gamma H_D m_z t,
\label{eqn: phi}
\end{equation}
where we now add a subscript to distinguish it from the AFM case.
Similar precessional dynamics can be derived for AFM. Substitution of $\mathbf{m}=(\mathbf{m}_1+\mathbf{m}_2)/2$ and the N\'eel order parameter $\mathbf{l}=(\mathbf{m}_1-\mathbf{m}_2)/2$ in the LLG equation gives $\dot{\mathbf{l}}=\mathbf{l} \times (\gamma H_\mathrm{ex} \mathbf{m}+\alpha \dot{\mathbf{m}})$ \cite{hals_phenomenology_2011}. In the low-damping limit we obtain $\dot{\mathbf{l}}=\gamma H_\mathrm{ex} (\mathbf{l} \times \mathbf{m})$. Due to large $H_{\rm ex}$, now $|\mathbf{m}| \ll 1$ and $\mathbf{l} \approx \mathbf{m}_1 \approx - \mathbf{m}_2$. The N\'eel vector $\mathbf{l}$ is restricted to the plane orthogonal to $\mathbf{m}$. If we define a cylindrical coordinate system where $\mathbf{m}$ is aligned along the $z$ axis, the  N\'eel vector $\mathbf{l}$ can be expressed by $l_x=\cos(\varphi_\mathrm{AFM})$, $l_y=\sin(\varphi_\mathrm{AFM})$ and $l_z = 0$. The time derivative of the angle of precession of $\mathbf{l}$ is $\dot{\varphi}_\mathrm{AFM}=- \gamma H_\mathrm{ex} |\mathbf{m}|$. By integrating, we obtain
\begin{equation}
\varphi_{\rm AFM}(t)=- \gamma H_\mathrm{ex} |\mathbf{m}| t.
\label{eqn: vecphiAFM}
\end{equation}

Comparing Eqns. \ref{eqn: phi} and \ref{eqn: vecphiAFM} we see that the low damping deterministic time dynamics for EPM and AFM are very similar. Both magnetizations precess around their intrinsic fields. The similarity in magnet dynamics between EPM and AFM is highlighted in Fig. \ref{fig: fig1}.

In a deterministic system, due to Gilbert damping, the perpendicular components of the magnetization ($m_z$ for EPM and $\mathbf{m}$ for AFM) eventually decrease to zero and the magnet becomes static. However, with the introduction of thermal noise, the perpendicular components fluctuate due to thermal torques so that $H_D$ and $H_\mathrm{ex}$ will lead to precession as described by Eqns. \ref{eqn: phi} and \ref{eqn: vecphiAFM}.  In an ensemble picture, the different precession frequencies result in the dephasing of the order parameter.
\begin{figure}[t]
\centering
{\includegraphics[trim={0 0 10 5},clip,width=1\linewidth]{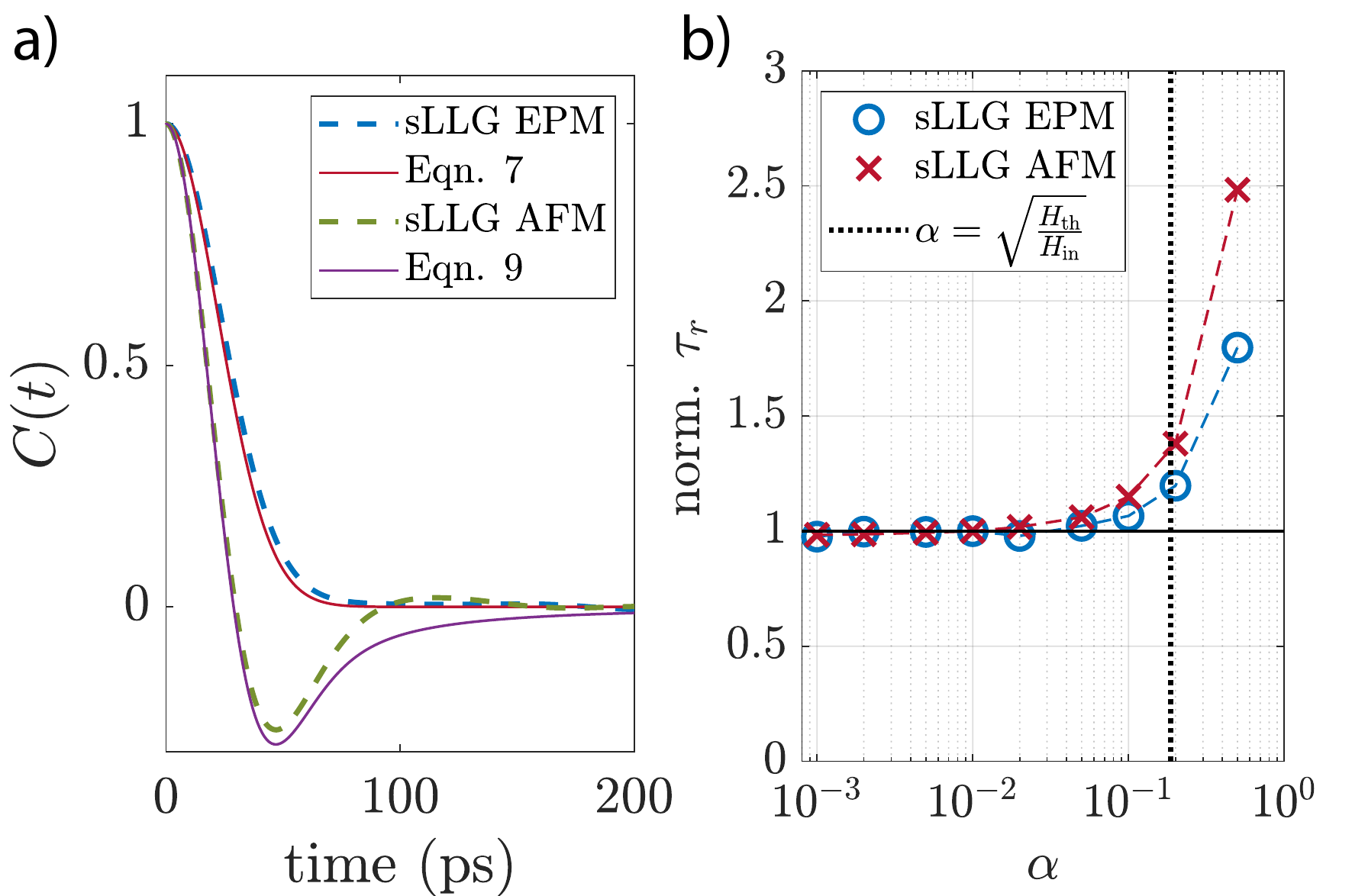}}
\caption{a) Normalized correlation function $C(t)$ of the order parameter $\mathcal{O}(t)$ for EPM and AFM in thermal equilibrium versus time. Numerical sLLG simulations are compared to the analytical Eqns. \ref{eqn: mx_c_EP} and \ref{eqn: mx_c_AFM}. b) Normalized average reversal time $\tau_r$ versus damping parameter $\alpha$. The average reversal time is extracted by setting $C(\tau_r)=1/2$ and normalized by $\tau_r(\alpha)/\tau_r(\alpha=0.01)$ for EPM and AFM.}
\label{fig: fig2}
\end{figure} 
We can make use of the fact that the perpendicular components in thermal equilibrium are distributed according to the Boltzmann distribution. With $\varphi_\mathrm{EPM}(t)$ given by Eqn. \ref{eqn: phi} and $m_x(t)=\cos(\varphi_\mathrm{EPM}(t))$, the normalized correlation function $C(t)=\langle \cos(\varphi_\mathrm{EPM}(t))\rangle$ for EPM where $\varphi_\mathrm{EPM}(0)=0$ can be evaluated by solving \citep{brown_thermal_1979,jaynes_information_1957}
\begin{equation}
C_\mathrm{EPM}(t)=\frac{\int_{-1}^{+1}dm_z \cos(\gamma H_D m_z t) \exp{[-\mathcal{F}_\mathrm{EPM} V/(k_B T)]}}{\int_{-1}^{+1} dm_z \exp{[-\mathcal{F}_\mathrm{EPM} V /(k_B T)]}}.
\end{equation} 
Due to the large demagnetization field, the integral has significant contributions only for small $|m_z|$. Thus,  we can extend the integration boundaries to $\pm \infty$. The integral evaluates to
\begin{equation}
C_\mathrm{EPM}(t)=\exp (-\omega_D^2  t^2/2),
\label{eqn: mx_c_EP}
\end{equation}
with $\omega_D=\gamma \sqrt{H_\mathrm{th}H_D}$.

For AFM we note that the plane of precession for $\mathbf{l}$ is not fixed to the $x-y$ plane. Instead, $\mathbf{l}$ rotates in a plane perpendicular to $\mathbf{m}$. For a given $\mathbf{l}$, $\mathbf{m}$ is in turn bounded to a plane perpendicular to $\mathbf{l}$. Hence, the Boltzmann integral for the normalized correlation function of $l_x(t)=\cos(\varphi_\mathrm{AFM}(t))$ becomes two-dimensional. Together with Eqn. \ref{eqn: vecphiAFM} we obtain
\begin{equation}
C_\mathrm{AFM}(t)= \frac{\int_{0}^{2\pi} \int_{0}^{1} d\rho d \theta \ \rho \cos{(\gamma H_\mathrm{ex} \rho t)} \exp{[-\mathcal{F}_\mathrm{AFM} V/(k_B T)]}}{\int_{0}^{2\pi} \int_0^{1} d\rho d \theta \ \rho \ \exp{[-\mathcal{F}_\mathrm{AFM} V/(k_B T)]}},
\label{eqn: c_AFM}
\end{equation}
where $\rho=\sqrt{m_y^2+m_z^2}$ with $m_y$ and $m_z$ being the $y$ and $z$ component of the total magnetization vector $\mathbf{m}$.
Solving the integrals with $\rho \rightarrow \infty$ gives
\begin{equation}
C_\mathrm{AFM}(t)= 1-\sqrt{2} \ \omega_\mathrm{ex} \ t \ D_F(\omega_\mathrm{ex} t/\sqrt{2}),
\label{eqn: mx_c_AFM}
\end{equation}
with $\omega_\mathrm{ex}=\gamma \sqrt{H_\mathrm{ex}H_\mathrm{th}}$ and the Dawson function defined by $D_F(x)=\exp{(-x^2) \int_0^x dt \exp(t^2)}$. 

Eqns. \ref{eqn: mx_c_EP} and \ref{eqn: mx_c_AFM} are plotted in Fig. \ref{fig: fig2} (a) as a function of time $t$ together with numerical sLLG solutions for the correlation function for EPM and AFM, respectively. In appendix \ref{app: num sim}, the numerical simulations are explained in detail. The following parameters are used for the numerical simulations throughout this Paper: $M_s= 1100 \mathrm{emu/cm^3}$, diameter $D=10$ nm, thickness $d_z= 1$ nm, $H_K= 1$ Oe, temperature $T= 300$ K, $H_{\rm th}=k_B T/(M_S V) = 479.4 \ \mathrm{Oe}$, $H_{\rm in}=H_D=H_{\rm ex}=4 \pi M_s=13.82 \ \mathrm{kOe}$. To extract the relevant time scale for the average reversal from the numerical simulations, instead of initializing an ensemble of nanomagnets in equilibrium, a single magnet is simulated over a long time sequence and the autocorrelation function is computed. The derived equations are in good agreement with the numerical simulations. Small deviations can be explained by the fact that the time sequence for calculating the autocorrelation is finite.

By setting $C(\tau_r)=1/2$ and solving for $\tau_r$ in Eqns. \ref{eqn: mx_c_EP} and \ref{eqn: mx_c_AFM}, we obtain a time scale for average reversal of the form $\tau_r \propto 1/(\gamma \sqrt{H_\mathrm{th} H_\mathrm{in}})$ which is Eqn. \ref{eqn: eqn_tau_r}. In comparison to the fundamental limit for the relaxation time of uniaxial anisotropy magnets in the low-barrier approximation $\tau_r \propto 1/(\alpha \gamma H_\mathrm{th})$, we obtain a speed up of the reversal time by a factor of $\sqrt{H_\mathrm{in}/(\alpha^2 H_\mathrm{th})}$ which is of the order of 2-3 magnitudes for typical experimental parameters at room temperature ($H_\mathrm{in} \approx 10^4$ Oe, $M_s=1100$ Oe, $V \approx 75 \ \mathrm{nm}^3$ and $\alpha \approx 0.01$; see appendix \ref{app: rev time} for more details).

Eqns. \ref{eqn: mx_c_EP} and \ref{eqn: mx_c_AFM} suggest a damping independent reversal time. In Fig. \ref{fig: fig2} (b) the normalized average reversal time is shown as a function of $\alpha$. For $\alpha \ll \sqrt{H_\mathrm{th}/H_\mathrm{in}}$, the average reversal time is independent of $\alpha$. This can be understood by noting that in this regime, the reversal time of the order parameter is described only by a precessional motion with average frequencies $\omega_D$ and $\omega_\mathrm{ex}$. The limit for this independence can be derived as follows: we assume assumed that $\dot{m}_z \approx 0$ (Eqn. \ref{eqn: dphidt}). However for EPM, we note that finite damping results in the change of $m_z$ on a characteristic time scale of $\tau_\mathrm{mz} \propto 1/(\alpha \gamma H_D)$. If $\tau_r$ is comparable to $\tau_\mathrm{mz}$, we can no longer assume that $\dot{m}_z \approx 0$.  The validity range of $\alpha$ can be obtained by using  $\tau_r^\mathrm{EPM} \ll \tau_\mathrm{mz}$, which is $\alpha   \ll \sqrt{H_\mathrm{th}/H_D}$ for EPM.  Using similar arguments for AFM in combination with the correspondence between $H_D$ and $H_{\rm ex}$, the general validity condition becomes  $\alpha \ll \sqrt{H_\mathrm{th}/H_\mathrm{in}}$. 

It has to be noted that for small damping, $1/(\alpha \gamma H_{\rm in}) \gg \tau_r$ which leads to precession of the order parameter (see Eqns. \ref{eqn: phi} and \ref{eqn: vecphiAFM}). As a result, the order parameter can still be correlated after its reversal. In the next section, we derive the time scale over which the order parameter becomes uncorrelated.

\section{MEMORY LOSS}
To answer the question of how long it takes for the magnetization to lose its memory, we initialize an ensemble of magnets with identical phase and perpendicular components. The system is now out of equilibrium and we cannot use Boltzmann law to obtain the correlation function. However, for EPM we can derive an analytical expression starting from the Langevin equation \cite{lemons_paul_1997,coffey_langevin_2012} for the angular coordinate $\varphi=\varphi_\mathrm{EPM}$
\begin{equation}
\Big[\tau_\mathrm{mz} \frac{d^2}{dt^2}+\frac{d}{dt}\Big] \varphi(t) =\Omega_\varphi (t),
\label{eqn: phid}
\end{equation}
where $\tau_\mathrm{mz}=(1+\alpha^2)/(\alpha \gamma H_D)$ and $\Omega_\varphi(t)$ is the stochastic fluctuation source term \footnote{Eqn. \ref{eqn: phid} can be derived from Eqn. \ref{eqn: dphidt} by eliminating $m_z$ and adding the stochastic fluctuation source term}. By applying Isserlis theorem the correlation of the $m_x$ component, $C_\mathrm{EPM}(t)=\langle m_x(t) m_x(0) \rangle= \langle \cos(\varphi) \rangle$  can be expressed as 
\begin{equation}
C_\mathrm{EPM}(t)=\exp \Big(-D_\varphi \Big[t-\frac{\tau_\mathrm{mz}}{2}(3+e^{-2t/\tau_\mathrm{mz}}-4e^{-t/\tau_\mathrm{mz}}) \Big] \Big),
\label{eqn: C_init}
\end{equation}
where $D_\varphi = \gamma H_\mathrm{th}/(\alpha (1+\alpha^2))$. The full derivation of Eqn. \ref{eqn: C_init} is given in appendix \ref{app: derivation}. This approach can also be used for the derivation of Eqn. \ref{eqn: mx_c_EP} of the previous section. For small Gilbert damping, Eqn. \ref{eqn: C_init} can be expressed as 
\begin{equation}
C_\mathrm{EPM}(t) \approx \exp (\alpha \gamma^3 H_D^2 H_\mathrm{th} t^3/3).
\label{eqn: C_init_sim}
\end{equation}
\begin{figure}[t]
\centering
{\includegraphics[trim={0 0 10 5},clip,width=1\linewidth]{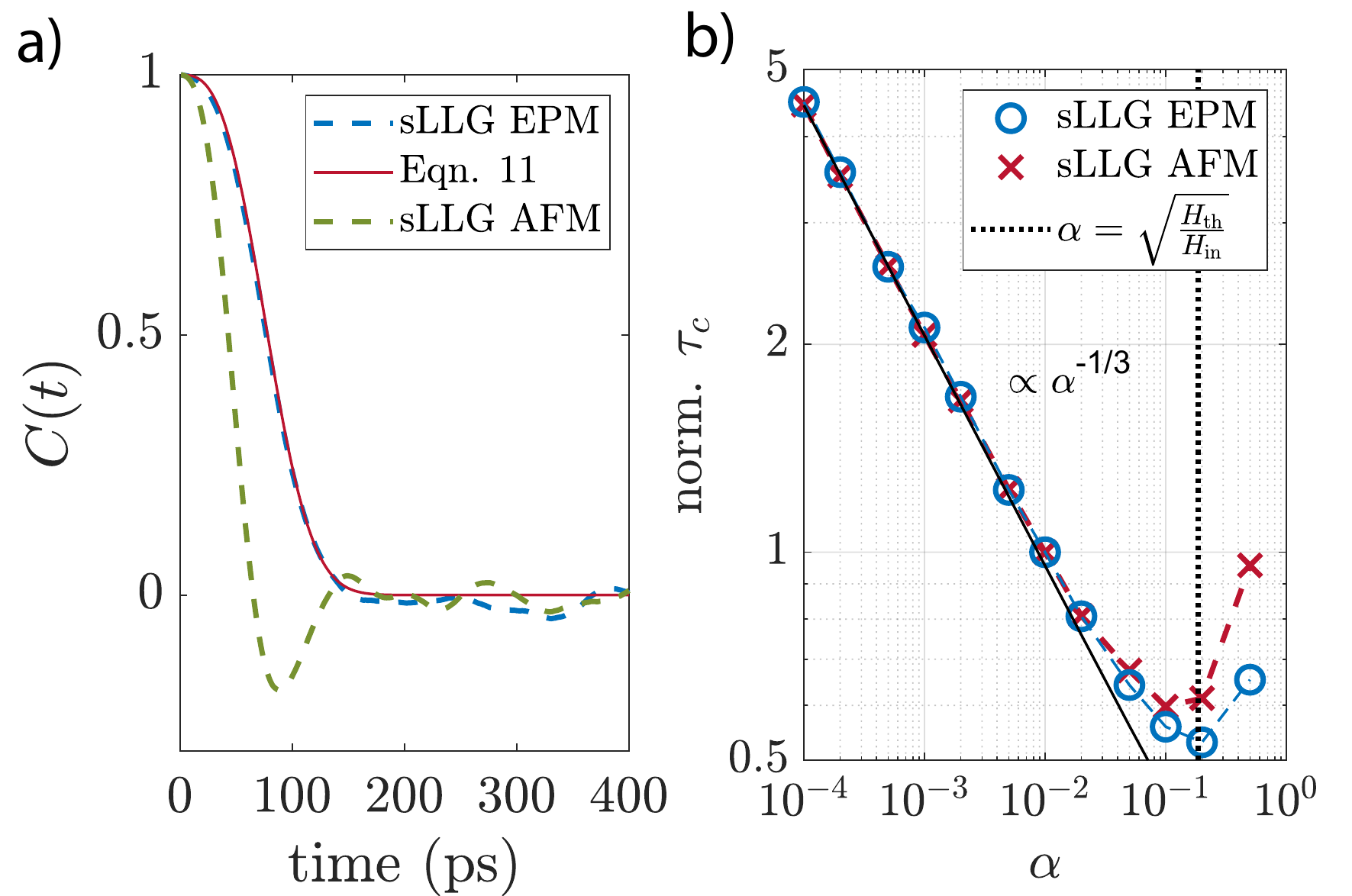}}
\caption{a) Normalized correlation function $C(t)$ for EPM and AFM with fixed initial conditions versus time $t$ . At $t=0$ $m_z$ of each magnet is set to $m_z=0$. b) Normalized correlation time $\tau_c$ versus damping parameter $\alpha$. The correlation time is extracted by setting $C(\tau_c)=1/2$ and normalized by $\tau_c(\alpha)/\tau_c(\alpha=0.01)$ for EPM and AFM. The numerical results are obtained by averaging over 1000 ensembles.}
\label{fig: fig3}
\end{figure} 
Eqn. \ref{eqn: C_init} is plotted in Fig. \ref{fig: fig3} (a) together with numerical results of the sLLG equation. 
Deriving nonequilibrium correlation functions for AFM requires generalizing the fluctuating dynamics, i.e. Eqn. \ref{eqn: phid}, to higher dimensions (as ${\bf l}$ can lie on a sphere) and requires further investigation. However, numerical results essentially show the same time scales and parameter dependencies for AFM when applying the correspondence of $H_D=H_\mathrm{ex}$.

By setting $C(\tau_c)=1/2$ in Eqn. \ref{eqn: C_init_sim} and solving for $\tau_c$, we obtain a time scale for memory loss. It scales as $1/(\alpha^{1/3} \gamma H_\mathrm{in}^{2/3} H_\mathrm{th}^{1/3})$ which is Eqn. \ref{eqn: eqn_tau_ensemble}. In comparison to the equation for the relaxation time for uniaxial anisotropy magnets in the low-barrier approximation $\tau_c \propto 1/(\alpha \gamma H_\mathrm{th})$, we find a speed up of memory loss by a factor of $H_\mathrm{in}/(\alpha H_\mathrm{th})^{2/3}$ which is of the order of 2 magnitudes for typical experimental values at room temperature ($H_\mathrm{in} \approx 10^4$ Oe, $M_s=1100$ Oe, $V \approx 75 \ \mathrm{nm}^3$ and $\alpha \approx 0.01$; see appendix \ref{app: mem loss} for more details).

Fig. \ref{fig: fig3} (b) shows the $\alpha$ dependence of the correlation time for AFM and EPM. The correlation time changes proportional to $\alpha^{-1/3}$ as predicted by Eqn. \ref{eqn: C_init_sim}. AFM shows the same dependence as EPM. These decorrelation dynamics stand in contrast to the memory loss of uniaxial anisotropy magnets exhibiting $\tau_c \propto \alpha^{-1}$. Similar to the reversal time, the regime for the $\alpha$ dependence of $\tau_c \propto \alpha^{-1/3}$ is limited by $\alpha \ll\sqrt{H_\mathrm{th}/H_\mathrm{in}}$. 

\section{DISCUSSION}
In this Paper, we show how intrinsic fields affect the dynamics of thermally excited magnetization of low-barrier EPM and AFM. This includes the theoretical understanding of the average reversal time as well as the memory-loss time of the order parameter. The time scales of the random fluctuations can be subnanosecond and 2-3 orders of magnitudes smaller than for nanomagnets with uniaxial anisotropy. This implies that EPM and AFM nanomagnets can be utilized to increase the operating speeds of applications that rely on large amounts of random numbers. This includes probabilistic computing, stochastic optimization, statistical sampling, cryptography and machine learning \cite{camsari_stochastic_2017,faria_low-barrier_2017,sutton_intrinsic_2017,merolla_million_2014,maass_noise_2014,bucci_high-speed_2003}.

The fluctuating order parameter can be read out by utilizing magnetic tunnel junctions (MTJs) \cite{julliere_tunneling_1975,*miyazaki_giant_1995,borders_integer_2019}. In MTJs, the fluctuating nanomagnets can be integrated as free layers that lead to fast resistance fluctuations \cite{camsari_implementing_2017,hassan_low-barrier_2019}. For low-barrier AFM, the free layer can be built as SAF structure consisting of two low-barrier magnets coupled by an exchange coupling layer. 

For the parameters chosen in this Paper in the case of EPM, typical demagnetizing fields lead to average reversal times of $\tau_r^\mathrm{EPM} \approx 25$ ps and correlation times of $\tau_c^\mathrm{EPM} \approx 75$ ps. For SAF free layers, high exchange interaction values of $J = 1- 30 \ \mathrm{erg / cm^2}$ \cite{parkin_systematic_1991,*zoll_giant_1997,*zoll_preserved_1997,*yakushiji_very_2017} lead to average reversal times of $\tau_r^\mathrm{AFM} \approx 6-32$ ps and correlation times of $\tau_c^\mathrm{AFM} \approx 10-100$ ps (compare Figs. \ref{fig: tau_r} and \ref{fig: tau_c}). Besides the reduced fluctuation time scales, SAF free layers have the advantage that the fixed layer in the MTJ can be a perpendicular anisotropy magnet (PMA) which is commonly used in MTJ technology \cite{ikeda_perpendicular-anisotropy_2010,park_systematic_2015}.

\begin{acknowledgments}
The authors thank Ernesto E. Marinero, Orchi Hassan and Rafatul Faria for insightful discussions. This work was supported in part by ASCENT, one of six centers in JUMP, a Semiconductor Research Corporation (SRC) program sponsored by DARPA. P.U. acknowledges support from the Purdue University Startup Funds.
\end{acknowledgments}

\renewcommand{\thetable}{A\arabic{table}}  
\renewcommand{\thefigure}{A\arabic{figure}} 
\renewcommand{\theequation}{A\arabic{equation}} 
\setcounter{figure}{0}    
\setcounter{equation}{0}   
\appendix
\section{Numerical simulations}
\label{app: num sim}
\begin{figure}[h]
\centering
{\includegraphics[width=0.7\linewidth]{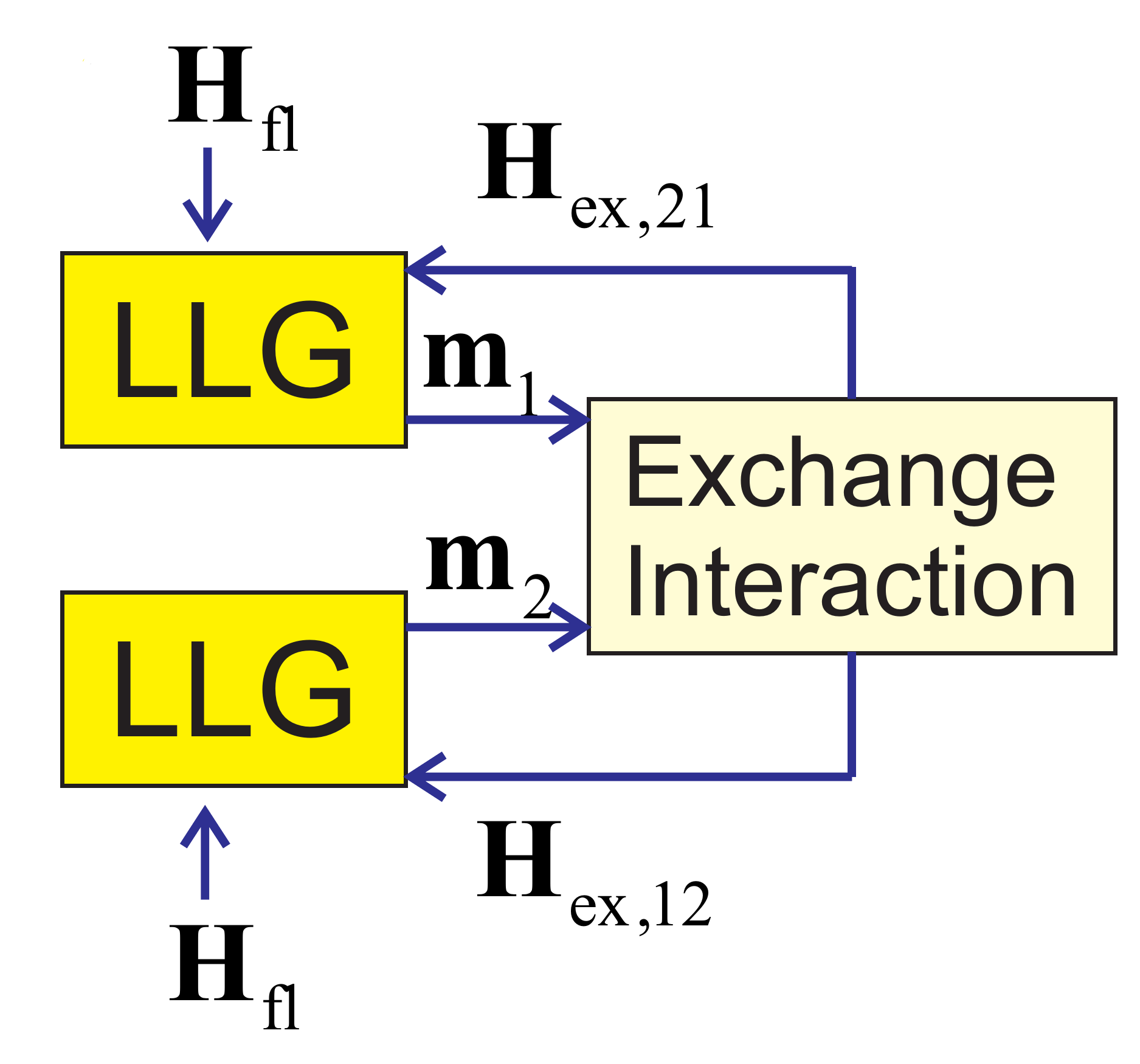}}

\caption{Coupled LLG modules for AFM simulations. Sublattices $\mathbf{m}_1$ and $\mathbf{m}_2$ are coupled with an exchange interaction module.}
\label{fig: LLG}
\end{figure}

In this section, the numerical simulations for solving the stochastic Landau-Liftshitz-Gilbert equations are described in detail. The simulations are performed with SPICE in the framework of the modular approach to spintronics \cite{camsari_modular_2015}. The sLLG equation for the time dynamics of magnet/sublattice $i$ is given by
\begin{equation}
(1+\alpha^2)\frac{d \mathbf{m}_i}{dt}=- \gamma \mathbf{m}_i \times \mathbf{H}_\mathrm{eff,i} - \alpha \gamma \mathbf{m}_i \times (\mathbf{m}_i \times \mathbf{H}_\mathrm{eff,i}),
\end{equation}
where $\mathbf{m}=\mathbf{M}/M_s$ is the magnetization unit vector with the saturation magnetization $M_s$, $\gamma$ is the gyromagnetic ratio, $\alpha$ is the damping constant and $\mathbf{H}_\mathrm{eff,i}=\mathbf{H}_K+\mathbf{H}_D+\mathbf{H}_\mathrm{ex,ij}+\mathbf{H}_\mathrm{fl}$ is the effective magnetic field. Here, $\mathbf{H}_K$ is the uniaxial anisotropy field, $\mathbf{H}_\mathrm{fl}$ is the field due to thermal fluctuations and $\mathbf{H}_\mathrm{ex,ij}=H_\mathrm{ex} \ \mathbf{m}_j$ where $i,j \in \{1,2\},i \neq j$ is the exchange field on $\mathbf{m}_i$ due to $\mathbf{m}_j$ being present only in AFM. For AFM, two LLG equations are coupled through an exchange interaction module as shown in Fig. \ref{fig: LLG}.
The thermal noise field $\mathbf{H}_\mathrm{fl}(t)$ is assumed to be Gaussian distributed with a zero mean. The standard deviation is given by the fluctuation dissipation theorem \cite{brown_thermal_1963,brown_thermal_1979}:
\begin{subequations}
\begin{gather}
\langle H_\mathrm{fl}^\mathrm{i}(t) \rangle =0,\\
\langle H_\mathrm{fl}^\mathrm{i}(t) H_\mathrm{fl}^\mathrm{j}(t')\rangle =\delta_\mathrm{ij} \delta(t-t') \sigma^2,\\
\sigma^2=\frac{2 \alpha k_\mathrm{B} T}{\gamma M_\mathrm{s}V}.
\end{gather}
\label{eqn: Hfl}
\end{subequations}

\renewcommand{\thetable}{B\arabic{table}}  
\renewcommand{\thefigure}{B\arabic{figure}} 
\renewcommand{\theequation}{B\arabic{equation}} 
\setcounter{figure}{0}    
\setcounter{equation}{0}   
\section{Derivation of the correlation functions of EPM using the stochastic equation}
\label{app: derivation}
The Langevin equation for the angular coordinate for EPM is
\begin{equation}
\begin{split}
\left[ \tau_\mathrm{mz}  \dfrac{d^2}{dt^2} +  \dfrac{d}{dt} \right] \varphi(t) &= \Omega_\varphi (t),
\end{split}
\end{equation}
where $ \Omega_\varphi (t)$ is the stochastic fluctuation source term. This fluctuating source has zero mean $\langle \Omega_\varphi (t) \rangle = 0$ and instantaneous covariance $\langle \Omega_\varphi (t) \Omega_\varphi (t^\prime) \rangle = 2 D_\varphi \delta(t-t^\prime)$ where $D_\varphi = \dfrac{\gamma H_\mathrm{th}}{\alpha (1+\alpha^2)}$. \\


The first-order equation for EPM can be solved subject to initial conditions $\{\varphi_0, \omega_0 \}$  at initial time $t=0$
\begin{widetext}
\begin{equation}
\begin{split}
\omega(t) &= \omega_0 \, e^{-t/\tau_\mathrm{mz}} + \dfrac{1}{\tau_\mathrm{mz}} \int_0^t dt' \, \Omega_\varphi (t') e^{-(t-t')/\tau_\mathrm{mz}} \\
\varphi(t) &= \varphi_0 + \int_0^t dt' \, \omega(t')= \varphi_0 + \omega_0 \tau_\mathrm{mz} [1-e^{-t/\tau_\mathrm{mz}}] + \int_0^t dt'' \, \Omega_\varphi (t'') [1-e^{-(t-t'')/\tau_\mathrm{mz}}],\\
\end{split}
\end{equation}
\end{widetext}
where $\tau_\mathrm{mz} = \dfrac{1+\alpha^2}{\alpha \gamma H_D}$.
\subsection{Average reversal time} 
For finding the average reversal time in thermal equilibrium, the initial angular velocity is provided by the thermal field $\langle\omega_0^2\rangle = D_\varphi/\tau_\mathrm{mz}$. To evaluate the correlation function, we proceed using the cumulant expansion relation or Isserlis theorem.

When applying the cumulant expansion to the correlation function $C(t)=\langle m_x (t) m_x(0) \rangle$ (note that $\langle \omega_0 \Omega_{\varphi}(t) \rangle = 0$),
\begin{widetext}
\begin{equation}
\begin{split}
 C(t) &= \exp\left[-\dfrac{1}{2} \langle\omega_0^2\rangle \tau_\mathrm{mz}^2 \left[ 1 -  e^{-t/\tau_\mathrm{mz}} \right]^2 -\dfrac{1}{2} \int_{0}^t dt_1 \int_{0}^t dt_2 \, \left[ 1- e^{-(t-t_1)/\tau_\mathrm{mz}}\right] \left[ 1- e^{-(t-t_2)/\tau_\mathrm{mz}}\right] \langle \Omega_\varphi (t_1)\Omega_\varphi (t_2)\rangle\right] \\
&= \exp\left[ -\dfrac{1}{2} \langle\omega_0^2\rangle \tau_\mathrm{mz}^2 \left[ 1 -  e^{-t/\tau_\mathrm{mz}} \right]^2-D_\varphi  \,  \int_{0}^t dt_1 \, \left[ 1- e^{-(t-t_1)/\tau_\mathrm{mz}}\right]^2\right]. \\
\end{split}
\end{equation}
\end{widetext}
Upon simplifying, we obtain
\begin{equation}
C(t) = \exp\left[- D_\varphi \,  \left[ t - \tau_\mathrm{mz} \left( 1 -  e^{-t/\tau_\mathrm{mz}}\right) \right] \right].
\end{equation}

For small $\alpha$, we find Eqn. \ref{eqn: mx_c_EP}
\begin{equation}
\begin{split}
C(t) &\approx  \exp\left[- \dfrac{D_\varphi}{2\tau_\mathrm{mz}} t^2 \right],\\
\end{split}
\end{equation}
where $\dfrac{D_\varphi}{\tau_\mathrm{mz}} \approx \gamma^2 H_D H_\mathrm{th} $.\\

\subsection{Memory loss} 
For finding the time for memory loss of a single magnet, we initialize an ensemble of easy-plane magnets identically. We fix $m_z$, thereby fixing the angular velocity at $t=0$ by the relation $\dot{\varphi} = -\omega_D m_z$. Thus, in the case where $m_z (t=0) = m$, $\omega_0 = \dot{\varphi} (t=0) = -\omega_D \, m$, we obtain
\begin{widetext}
\begin{equation}
\begin{split}
C(t) &= \langle \cos\varphi (t) \rangle =\text{Re} \left[ \langle e^{i \varphi(t)} \rangle\right] \\
&=\dfrac{1}{2} \exp\left[ -i\omega_D m \tau_\mathrm{mz} ( 1 -  e^{-t/\tau_\mathrm{mz}} )\right] \langle \exp\left[ i \int_{0}^t dt^\prime \, \left[ 1- e^{-(t-t^\prime)/\tau_\mathrm{mz}}\right] \Omega_\varphi (t^\prime)\right] \rangle \\
&+ \dfrac{1}{2} \exp\left[ i\omega_D m \tau_\mathrm{mz} ( 1 -  e^{-t/\tau_\mathrm{mz}} )\right] \langle \exp\left[- i \int_{0}^t dt^\prime \, \left[ 1- e^{-(t-t^\prime)/\tau_\mathrm{mz}}\right] \Omega_\varphi (t^\prime)\right] \rangle.
\end{split}
\end{equation}
\end{widetext}
Applying cumulant expansion or Isserlis theorem to evaluate the stochastic average with $m=0$, we obtain
\begin{equation}
C(t) = \exp\left[ - D_\varphi \,  \left[ t - \dfrac{\tau_\mathrm{mz}}{2} \left( 3 + e^{-2t/\tau_\mathrm{mz}} - 4 e^{-t/\tau_\mathrm{mz}}\right)\right]\right],
\label{eqn: Cf}
\end{equation}
which is Eqn. \ref{eqn: mx_c_AFM}.

\renewcommand{\thetable}{C\arabic{table}}  
\renewcommand{\thefigure}{C\arabic{figure}} 
\renewcommand{\theequation}{C\arabic{equation}} 
\setcounter{figure}{0}    
\setcounter{equation}{0}  

\section{Average reversal time for different intrinsic fields}
\label{app: rev time}
\begin{figure}[t!]
\centering
{\includegraphics[trim={0 4 25 0},clip,width=1\linewidth]{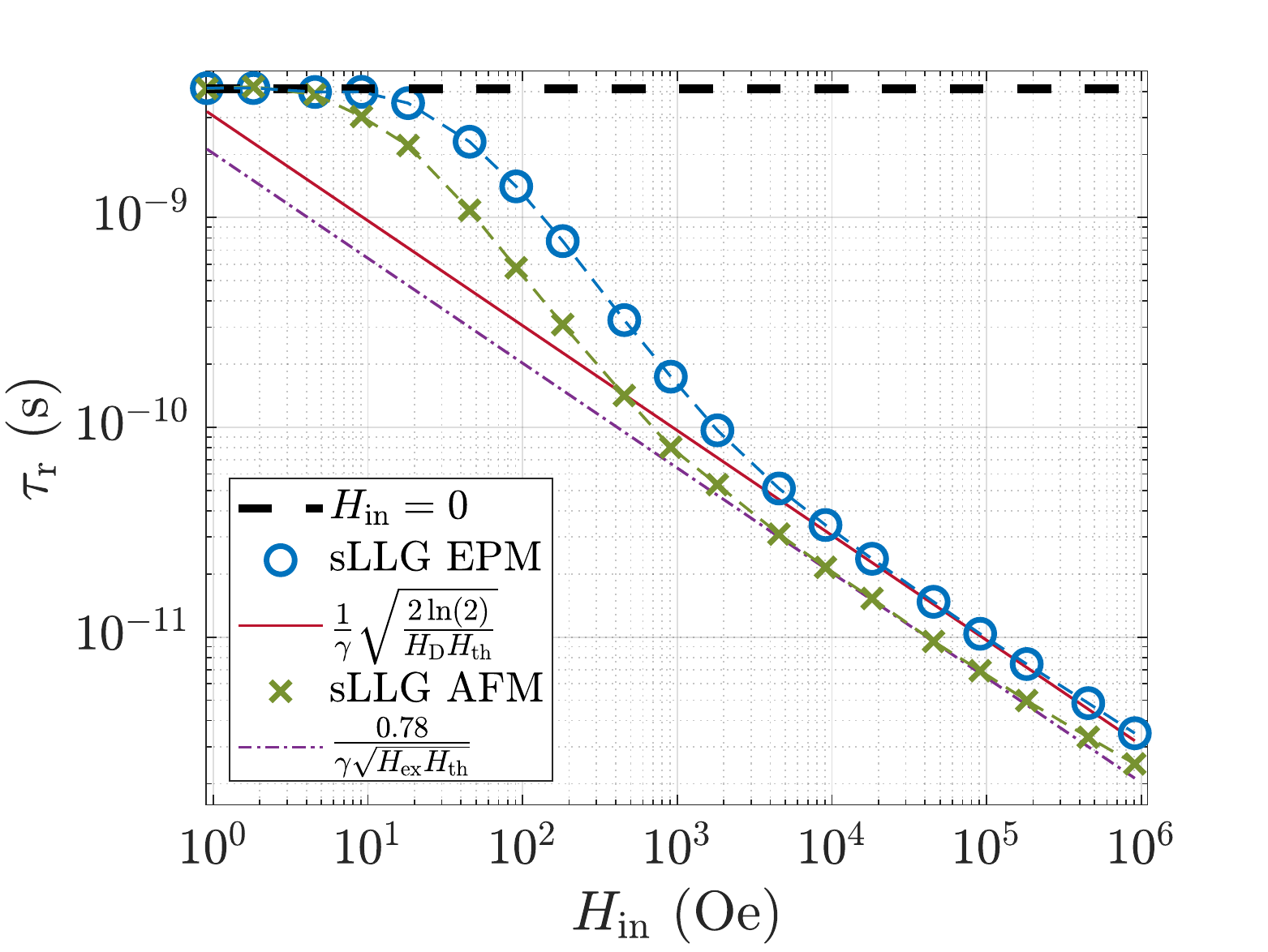}}

\caption{Average reversal time $\tau_r$ versus $H_\mathrm{in}$ ($H_D$ for EPM and $H_\mathrm{ex}$ for AFM). The following parameters are used in the numerical simulations: saturation magnetization $M_s=1100 \ \mathrm{emu/cm^3}$, magnet diameter $D=10 \ \mathrm{nm}$, thickness $d_z=1 \ \mathrm{nm}$, $H_K=1 \ \mathrm{Oe}$, $\alpha=0.01$.}
\label{fig: tau_r}
\end{figure}

Eqns. \ref{eqn: mx_c_EP} and \ref{eqn: mx_c_AFM} are the normalized correlation functions for the average reversal of EPM and AFM in thermal equilibrium. By setting $C(\tau_r)=1/2$ and solving for $\tau_r$ we can extract a time scale for the reversal time. For EPM, we obtain $\tau_r^{EPM}=\sqrt{2 \ln(2)/(\gamma^2 H_D H_\mathrm{th})}$. For AFM, Eqn. \ref{eqn: mx_c_AFM} is not directly invertable. However, numerically a prefactor of $0.78$ can be obtained so that $\tau_r^{AFM}=0.78/(\gamma \sqrt{H_\mathrm{ex} H_\mathrm{th}})$. In Fig. \ref{fig: tau_r} the numerically and analytically reversal time for EPM and AFM are shown and compared to the fundamental limit of the reversal time of uniaxial anisotropy low-barrier magnets, which is given by $\ln(2)/(2 \alpha \gamma H_\mathrm{th})$. In the experimental relevant regime, the reversal time is in the subnanosecond regime and about 2-3 orders of magnitude smaller than the reversal time of uniaxial anisotropy magnets, where $H_\mathrm{in}=0$.

 \renewcommand{\thetable}{D\arabic{table}}  
\renewcommand{\thefigure}{D\arabic{figure}} 
\renewcommand{\theequation}{D\arabic{equation}} 
\setcounter{figure}{0}    
\setcounter{equation}{0}  
\section{Memory loss for different intrinsic fields} 
\label{app: mem loss}
\begin{figure}[t!]
\centering
{\includegraphics[trim={0 4 25 0},clip,width=1\linewidth]{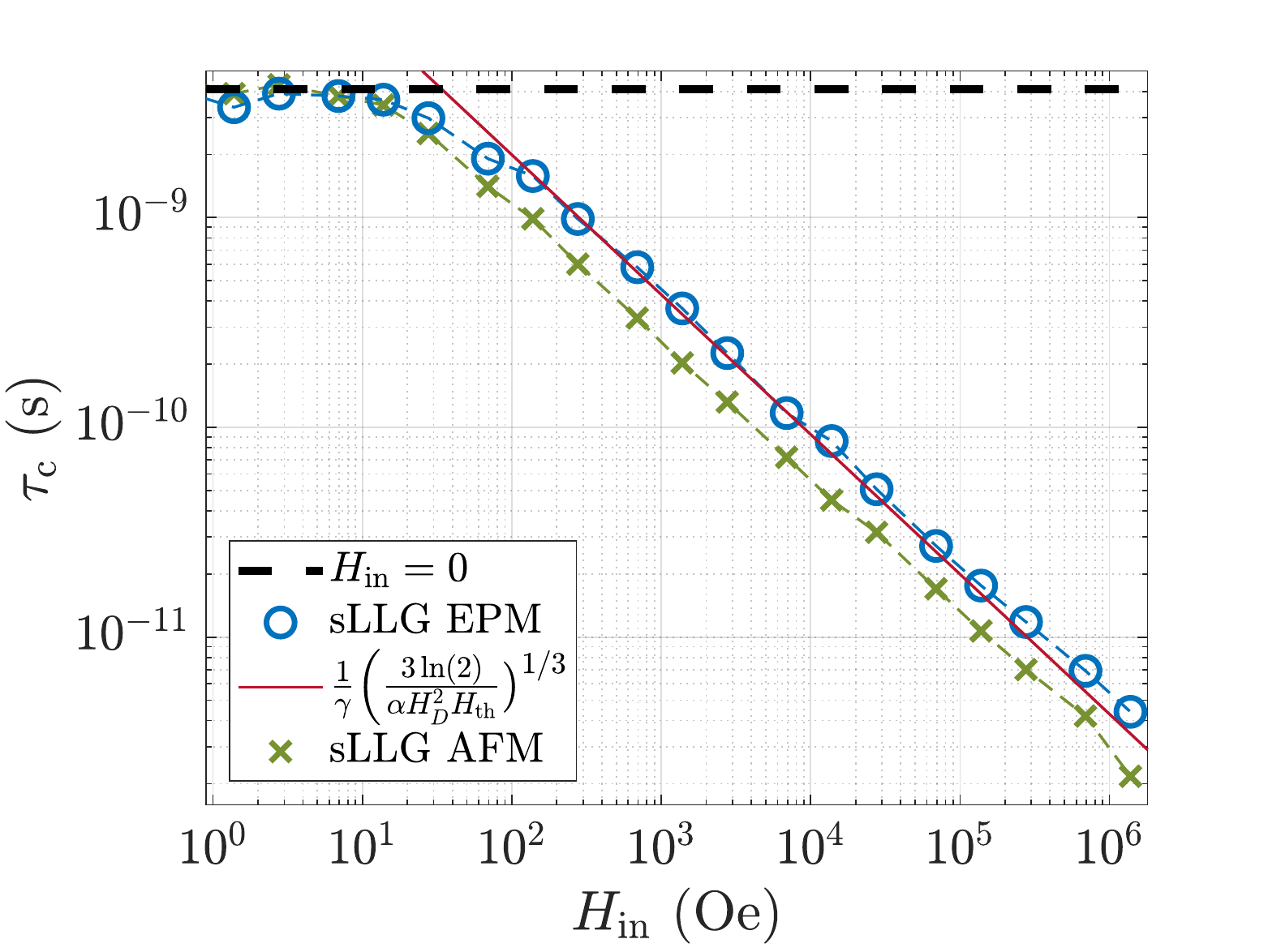}}

\caption{Correlation time $\tau_c$ versus $H_\mathrm{in}$ ($H_D$ for EPM and $H_\mathrm{ex}$ for AFM). The following parameters are used in the numerical simulations: saturation magnetization $M_s=1100 \ \mathrm{emu/cc}$, magnet diameter $D=10 \ \mathrm{nm}$, thickness $d_z=1 \ \mathrm{nm}$, $H_K=1 \ \mathrm{Oe}$, $\alpha=0.01$. The numerical results are obtained by averaging the order parameter over 100 ensembles.}
\label{fig: tau_c}
\end{figure}

Eqn. \ref{eqn: C_init} is the correlation function for an ensemble of EPMs initialized at $m_z=0$. By setting $C(\tau_c)=1/2$ and solving for $\tau_c$ we can extract a time scale for memory loss which evaluates to $\tau_c^{EPM}=[3 \ln(2)/(\alpha \gamma^3 H_D^2 H_\mathrm{th})]^{1/3}$. In Fig. \ref{fig: tau_c}, the time for memory loss for EPM and AFM are shown and compared to the fundamental limit of the relaxation time of uniaxial anisotropy low-barrier magnets given by $\ln(2)/(2 \alpha \gamma H_\mathrm{th})$. For AFM, no analytic equation was derived. However, similar to the reversal time, the correlation time is about a factor of 2/3 smaller for AFM than for EPM. In the experimental relevant regime, the correlation time is in the subnanosecond regime and about 2 orders of magnitude smaller than the relaxation time of uniaxial anisotropy magnets, where $H_\mathrm{in}=0$.

\bibliographystyle{apsrev4-1}
\bibliography{library}
\end{document}